\documentclass[conference]{IEEEtran}
\IEEEoverridecommandlockouts
\usepackage{cite}
\usepackage{amsmath,amssymb,amsfonts}
\usepackage{algorithmic}
\usepackage{graphicx}
\usepackage{textcomp}
\usepackage{xcolor}
\usepackage{float}
\usepackage{verbatimbox}

\usepackage{listings}
\usepackage{color}
\usepackage{subfig}

\def\BibTeX{{\rm B\kern-.05em{\sc i\kern-.025em b}\kern-.08em
    T\kern-.1667em\lower.7ex\hbox{E}\kern-.125emX}}
    
\newcommand{\ket}[1]{\left | \, #1 \right\rangle}

\newcommand{\bea}{\begin{eqnarray}}
\newcommand{\eea}{\end{eqnarray}}
    
\begin{document}

\title{Procedural generation using quantum computation}

\author{
\IEEEauthorblockN{James R. Wootton}
\IEEEauthorblockA{
\textit{IBM Quantum, IBM Research - Zurich}\\
Switzerland \\
jwo@zurich.ibm.com
}
}

\maketitle

\begin{abstract}

Quantum computation is an emerging technology that promises to be a powerful tool in many areas. Though some years likely still remain until significant quantum advantage is demonstrated, the development of the technology has led to a range of valuable resources. These include publicly available prototype quantum hardware, advanced simulators for small quantum programs and programming frameworks to test and develop quantum software. In this provocation paper we seek to demonstrate that these resources are sufficient to provide the first useful results in the field of procedural generation. This is done by introducing a proof-of-principle method: a quantum generalization of a blurring process, in which quantum interference is used to provide a unique effect. Through this we hope to show that further developments in the technology are not required before it becomes useful for procedural generation. Rather, fruitful experimentation with this new technology can begin now.

\end{abstract}

Quantum computation is a new technology based on a radically new form of hardware and software~\cite{benioff,feynman,ike-mike}. This will allow certain problems to be solved with a significant reduction in computational complexity in comparison with conventional digital computing~\cite{deutsch,montanaro}. The resulting `quantum speedup' varies from polynomial to super-polynomial or even exponential for algorithms address a variety of different types of problem~\cite{algorithm-zoo,qiskit-textbook}.

The basic unit of quantum computation is the qubit. To run most of the algorithms developed over the past few decades, thousands of qubits will be required. Quantum processors of this size do not currently exist, and are still some years away. At the other extreme, any small scale quantum process of up to around 20 qubits can easily be simulated by a laptop, and up to around 50 qubits can be simulated by a supercomputer~\cite{google-supremacy,ibm-summit}.

Currently, most prototype quantum devices have no more than 20 qubits, and it is devices of up to this size that have been made publicly available through cloud services~\cite{ibm-nisq}. These have found a great deal of use by those seeking to test and understand quantum processors, such as those using them as experimental hardware to perform studies relating to the science behind the devices. However, as pieces of computational hardware, the fact of their easy simulability by conventional computers means that they cannot yet provide an advantage over conventional hardware.

Nevertheless, by considering the methodology required for quantum computing, we can start to think about how quantum algorithms can be used in different fields. For proof-of-principle cases of up to 20 qubits, we can easily implement these methods either by simulation or using prototype devices. For the most part, the results of these proof-of-principle implementations are not useful in themselves. However, in some cases we may find that they do indeed offer a unique insight that might not have been considered outside the context of quantum computation. Such a case would not represent an advantage of using quantum hardware over conventional hardware, since it is just as easy to use conventional hardware in a simulation. Instead it would represent an advantage of designing software using quantum principles.

In this paper, we suggest that procedural generation is one of the most likely fields in which such initial advantages might be found. The popular `wave function collapse' algorithm could be thought of as foreshadowing this~\cite{wfc}, since it is a quantum inspired method that has proven very useful to many practitioners of procedural generation. However, it is not an example in itself since it was not designed using the principles of quantum computing software.

To provide a more substantial motivating example, we must focus on finding applications for the kind of results that quantum computers find easiest to produce. These include simulations of quantum dynamics, and the generation of quantum interference effects. An example of how the former might be used in procedural generation can be found in \cite{q_avrai}, where it was shown that quantum simulations can provide a rudimentary AI for a \emph{Civilization}-like game. For the latter we can look to ideas such as quantum walks~\cite{kendon:20}, which provide a generalization of familiar random walks. It is a simple example of such a process that we will consider below, as a unique means of providing a blur effect.

\section{Introduction to quantum computation}

Before we consider an application of quantum computing, it is first necessary to explain the basic principles of quantum software. In this we will focus on the aspects that are most relevant for the application that we will consider.

As with standard digital computers, we begin with the concept of the `bit'. These are objects that are limited to two possible states, often denoted \texttt{0} and \texttt{1}. Most forms of hardware used to encode this information are governed by the laws of classical physics, which means that the bit value is definitely either \texttt{0} or \texttt{1}. However, it is also possible to create hardware in which the bit value is governed by the laws of quantum physics. To describe this, it is useful to associate the bit values with a pair of orthogonal vectors, written as $\ket{0}$ and $\ket{1}$. 

When we read out the bit value from such a quantum bit or `qubit', the states $\ket{0}$ and $\ket{1}$ will behave as one would expect: from $\ket{0}$ we read out \texttt{0}, and from $\ket{1}$ we read out \texttt{1}. However, quantum physics also allows states of the form
\[
c_0 \ket{0} + c_1 \ket{1},
\]
where the so-called `amplitudes' $c_0$ and $c_1$ are arbitrary complex numbers that satisfy $|c_0|^2 +|c_1|^2  = 1$. Such states are known as `superposition states'.

Reading out the bit value from a qubit in a superposition state will still result in a simple \texttt{0} or \texttt{1}. However, the output will be random in this case. Specifically, \texttt{0} will occur with probability $|c_0|^2$ and \texttt{1} with probability $|c_1|^2$. Once the bit value has been read out, the superposition state `collapses' to the corresponding state $\ket{0}$ or $\ket{1}$. For example, after obtaining the output \texttt{0}, the qubit will no longer be in a superposition state but instead will be in the simple state $\ket{0}$. This effect means that keeping track of the bit value is not a passive process, as it is in the classical case. Instead it has the ability to change the state of the qubit. As such, we must be careful to specify exactly when we want such readout events, known as measurements, to occur.

Superposition states are more than simply a source of randomness. They allow a wider variety of methods to manipulate the bit. Classically, there is a very limited set of operations that can be performed on just a single bit: set it to \texttt{0}, set it to \texttt{1} or flip the value with a \texttt{NOT} gate. In the quantum case, however, we can perform any of an infinite set of parameterised operations. All of these can be generated by the operations $R_x (\theta)$ and $R_y (\theta) $, which have the following effect.

\bea \nonumber
R_x (\theta)  \ket{0} &=&  i \cos \frac{\theta}{2}  \ket{0} + \sin \frac{\theta}{2} \ket{1} ,\\ \nonumber
R_x (\theta) \ket{1} &=&  \sin \frac{\theta}{2}  \ket{0} + i \cos \frac{\theta}{2}  \ket{1}, \\ \nonumber
R_y (\theta)  \ket{0} &=&  \cos \frac{\theta}{2}  \ket{0} + \sin \frac{\theta}{2} \ket{1}, \\ \nonumber
R_y (\theta) \ket{1} &=&  -\sin \frac{\theta}{2}  \ket{0} + \cos \frac{\theta}{2}  \ket{1}.
\eea
\vspace{0.25cm}

Note that $R_x(\pi)$ has the effect $\ket{0} \leftrightarrow \ket{1}$. The $R_x$ operation therefore reproduces and generalizes the effect of the \texttt{NOT} gate. Similarly, $R_y(\pi/2)$ has the effect $\ket{0} \rightarrow \ket{1}$ and $\ket{1} \rightarrow -\ket{0}$. The factor of $-1$ in the latter has no effect on the probabilities for the results of a measurement, and so $R_y$ also reproduces and generalizes the effect of the \texttt{NOT} gate, but in a different way to that done by $R_x$.

We have now seen that the quantum implementation of bits allows a greater variety of possible manipulations. However, we have not yet seen any example of why this may be advantageous. All such examples require the use of more than just one qubit. For two qubits, the possible states take the form
\[
c_{00} \ket{00} + c_{10} \ket{10}  + c_{01} \ket{01}  + c_{11} \ket{11} 
\]
The probability of obtaining the outcome described by each two bit string $\mathtt{ b_1 b_0} $  is $|c_{ \mathtt{ b_1 b_0} }|^2$, and the complex numbers $c_{ \mathtt{ b_1 b_0} }$ must obey the restriction $\sum_{\mathtt{ b_1 b_0}} |c_{ \mathtt{ b_1 b_0} }|^2 = 1$. For $n$ qubits the states take a corresponding form using $n$-bit strings.

So far we have introduced only single qubit manipulations. These must be supplemented with multi-qubit operations in order to perform computation. The most important of these is the so-called `controlled-\texttt{NOT}' or \texttt{cx} gate, which acts on a given pair of qubits. The effect is not symmetric, and so one qubit is designated the `control', and the other is the `target'. The effect is as follows, with the control in these examples written on the left.

\bea \nonumber
\texttt{cx} \ket{00} = \ket{00}, && \texttt{cx} \ket{01} = \ket{01}, \\ \nonumber
\texttt{cx} \ket{10} = \ket{11}, && \texttt{cx} \ket{11} = \ket{10}.
\eea
\vspace{0.25cm}

We can think of this operation as performing a $\texttt{NOT}$ on the target qubit iff the control qubit is in state $\ket{1}$. Alternatively, we can think of it as overwriting the target qubit state with the \texttt{XOR} of the inputs. In this sense it is a quantum (and reversible) implementation of the classical \texttt{XOR} gate.

Given just the ability to perform the $R_x$, $R_y$ and \texttt{cx} gates on all qubits, it is possible to transform any multi-qubit state into any other. In fact, it is possible to implement any mapping from a set of input bit strings to output bit strings, and so to reproduce any classical computation. However, this does not just provide an alternative form of hardware on which to implement the same algorithms as for digital computers. Instead, it provides a unique way of manipulating information~\footnote{As a quick plausibility argument for this, note that a \texttt{NAND} gate cannot be constructed in the Boolean circuit model using \texttt{NOT} and \texttt{XOR} gates alone. However, this can be achieved in the quantum circuit model using only the quantum generalizations of the \texttt{NOT} and \texttt{XOR} gates. This is a simple example of a computational task that quantum computaters can achieve in a way that standard digital computers cannot.}, leading to a variety of unique algorithms that can be implemented with quantum computation~\cite{qiskit-textbook}.

By considering the `textbook' quantum algorithms developed over the past few decades~\cite{algorithm-zoo}, we can begin to speculate on how they will be used in procedural generation. In particular the reduced computational complexity for constraint satisfiability problems could be useful in searching the probability space (for example~\cite{smith:10,smith:11}), and speed-ups for graph-theoretic analysis could help to find useful properties of problems expressed as networks (for example~\cite{kybartas:14}). However, the usefulness of these algorithms will depend on exactly what time and resources are required~\cite{campbell:19}. This will not be fully known until the scalable and fault-tolerant quantum hardware required to run these algorithms has been built, which is still some years away. However, these potential opportunities show that quantum computation will be useful for procedural generation in the long-term, which provides motivation to explore applications in the near-term.

\section{Encoding images in quantum states}

Using the concepts introduced in the last section, we will introduce a method to encode and manipulate images using multiqubit states. Specifically, the manipulation will be to implement a blur like effect using quantum interference. We will consider grayscale images, in which the brightness of a pixel corresponds to a value from $0$ to $1$. We can equivalently consider these images to be height maps, and will sometimes refer to them as such.

Code snippets are provided to show explicitly how the method can be generated. These are in Python, and will use the Qiskit framework for handling quantum circuits~\cite{qiskit}. First we begin by importing the necessary tools.

\vspace{0.25cm}
\begin{verbnobox}[\fontsize{8pt}{8pt}\selectfont]
import numpy as np
from math import pi
from qiskit import QuantumCircuit, quantum_info as qi
\end{verbnobox}
\vspace{0.25cm}

For other languages, note that MicroQiskit could be used~\cite{microqiskit}. This is a minimal reimplementation of Qiskit, designed to facilitate ports to other programming languages. The following methods can be used in MicroQiskit with almost identical syntax.

\subsection{Converting images to quantum states}

Though we are focussing on small-scale quantum processes of up to around 20 qubits, we will want to use them to generate images with thousands of points. Clearly there is a difference of scale between the two. However, as we saw above, the state of $n$ qubits is described by a set of $2^n$ amplitudes: one for each possible output bit string. We can therefore close the gap by making use of all of these amplitudes, or alternatively the corresponding probabilities.

When using a real quantum processor, the state cannot be accessed directly. Instead we need to estimate the probabilities for each of the $2^n$ possible output bit strings. This can be done by repeating a circuit many times to sample from the output. Specifically, using $\mathtt{shots}=4^n$ samples should be adequate to estimate the probabilities with sufficient accuracy. However, this number of samples carries a computational complexity that is greater than the $O(2^n)$ required to directly access the probabilities when a simulation is used. The method we will develop is therefore one for which the use of a simulator is in fact advantageous over the use of a real quantum device: the reverse of what is normally found in quantum computing.

Our first task is to find a mapping between the numbers that describe an image (brightness values for each coordinate) and the numbers that describe a quantum state (amplitudes for each output bit string). The most important element of this is to define a mapping between the coordinates and the bit strings.

The ideal mapping for our purposes would be one that maps neighbouring coordinates to neighbouring bit strings. For example, if we map some $(x,y)$ to \texttt{0000}, a good choice for neighbouring points would be

\bea \nonumber
(x+1,y) \, &\rightarrow& \mathtt{1000}   \\ \nonumber
(x-1,y) \, &\rightarrow& \mathtt{0100} ,  \\ \nonumber
(x,y+1) \, &\rightarrow& \mathtt{0010} , \\ \nonumber
(x,y-1) \, &\rightarrow& \mathtt{0001}.
\eea

Here the Manhattan distance between any two points is equal to the Hamming distance between the corresponding bit strings.

In general, this will not be a perfect mapping. We usually consider images based on 2D square lattices, whereas the structure formed by the Hamming distance between $n$-bit strings is an $n$-dimensional hypercube. This will mean that there will always have to be non-neighbouring coordinates (on the lattice) whose bit strings are neighbours (in the hypercube).  However, we can at least ensure that neighbouring coordinates always have neighbouring bit strings.

This is done using repeated applications of a process in which we take such a list for $n$-bit strings and use it to create a list with doubled length for $(n+1)$-bit strings. This process starts by creating two altered copies of the original list. For the first copy, we simply add a \texttt{0} to the end of each bit string. For the second we reverse the order and then add a \texttt{1} to the end of each bit string. Finally, these two lists are concatenated.

For example, starting with $\mathtt {\left[0,1 \right]} $ we get $\mathtt {\left[00,10 \right]} $ for the original list with appended \texttt{0}, and $\mathtt {\left[11,01 \right]} $ for the reversed list with appended \texttt{1}. These combine to form $\mathtt {\left[00,10,11,01 \right]} $. By repeating the process, we can obtain a list of length $2^n$ for $n$-bit strings for any desired $n$.

The process can be applied by the following Python function. Given a desired length for the list, this returns a list of at least that length.
 
\vspace{0.25cm}
\begin{verbnobox}[\fontsize{8pt}{8pt}\selectfont]
def make_line ( length ):
    n = int(np.ceil(np.log(length)/np.log(2)))
    line = ['0','1']
    for j in range(n-1):
        cp0 = []
        for string in line:
            cp0.append (string+'0')
        cp1 = []
        for string in line[::-1]:   
            cp1.append (string+'1')
        line = cp0+cp1
    return line
\end{verbnobox}
\vspace{0.25cm}

With this, we can combine the $x$th and $y$th elements of two such lists to define a unique string for each coordinate $(x,y)$ of a grid. The following function, \texttt{make\_grid} runs through all the coordinates of an $L\times L$ grid, calculates the corresponding bit string, and then outputs all the results. This is done as a Python dictionary, with bit strings as keys and the corresponding coordinates as values.

\vspace{0.25cm}
\begin{verbnobox}[\fontsize{8pt}{8pt}\selectfont]
def make_grid(L):
    line = make_line( L )
    grid = {}
    for x in range(L):
        for y in range(L):
            grid[ line[x]+line[y] ] = (x,y)
    return grid
\end{verbnobox}}
\vspace{0.25cm}

Now we have figured out what to do with the coordinates in an image, it is time to focus on the brightness values themselves. To do this, we will assume that each value $h$ exists in the range $0\leq h \leq 1$, and that the largest of all the heights is equal to exactly $1$. This assumption is without loss of generality, since any set of values can be shifted and rescaled into this form.

We then define a quantum state for which the probability of a bit string $b$ is proportional to the brightness of the corresponding point ${\tt grid[ } b {\tt \\]}$,

\[
\frac{ p_{b'} }{ p_{b} } = \frac{ h_{ {\tt grid[ } b' {\tt \\]} } }{ h_{ {\tt grid[} b {\tt \\]} } }
\]

The reason why we cannot simply set $p_{b} = h_{ {\tt grid[} b {\tt \\]} }$ is because probabilities must always sum to 1. To achieve this we simply renormalize using 

\[
p_{b} = \frac{ h_{ {\tt grid[} b {\tt \\]} } }{ H }, \,\,\,\, H = \sum_b h_{ {\tt grid[} b {\tt \\]} }.
\]

Now we have the probabilities, we need corresponding amplitudes for the basis states. When we restrict to the case that these amplitudes are real numbers, they are related to the probability by the simple relation $c_b = \sqrt{p_b}$. The state we require to encode our image is then

\[
\frac{1}{\sqrt{H}} \sum_b \sqrt{h_b} \, | b \rangle
\]

Now we can construct a function to create this state for any given image. This means that we need to create a `quantum circuit', a set of operations for a suitable number of qubits which will prepare the required state from the default initial $|00\ldots00\rangle$ state. Rather than dwell on the details of this at an abstract level, we can simply use Qiskit in which quantum circuits are defined within a \texttt{QuantumCircuit} class, which includes an \texttt{initialize} method that can be used to set desired initial states.

The images that we will manipulate in the following will be expressed in the form of Python dictionaries, with coordinates as keys and the corresponding brightness as values. Absent coordinates are assumed to correspond to a value of 0. Here is an example of such an image.

\vspace{0.25cm}
\begin{verbnobox}[\fontsize{8pt}{8pt}\selectfont]
height = {}
for pos in [(2,5),(2,6),(5,6),(5,5),(2,1),\
            (3,1),(4,1),(5,1),(1,2),(6,2)]:
    height[pos] = 1
\end{verbnobox}
\vspace{0.25cm}

The following function allows us to convert a given image to a quantum circuit. 

\vspace{0.25cm}
\begin{verbnobox}[\fontsize{8pt}{8pt}\selectfont]
def height2circuit(height):
    # determine grid size
    L = max(max(height))+1
    # make grid
    grid = make_grid(L)
    # determine required qubit number
    n = 2*int(np.ceil(np.log(L)/np.log(2)))
    # create empty state
    state = [0]*(2**n)
    # fill state with required amplitudes
    H = 0
    for bit string in grid:
        (x,y) = grid[bit string]
        if (x,y) in height:
            h = height[x,y]
            state[ int(bit string,2) ] = np.sqrt( h )
            H += h
    # normalize state
    for j,amp in enumerate(state):
        state[ j ] = amp/np.sqrt(H)
    # define and initialize quantum circuit            
    qc = QuantumCircuit(n,n)
    qc.initialize(state,range(n))
    # for standard Qiskit, use
    # qc.initialize( state, qc.qregs[0]) 
    return qc
\end{verbnobox}
\vspace{0.25cm}

\subsection{Converting quantum states to images}

The next job is to implement the opposite process: to turn a quantum circuit into an image. For this we must determine the probability  $p_{b}$ for each possible output bit string $b$. In the case that we are simulating the process we can keep track of the entire quantum state. The probabilities $p_{b}$ can then be extracted directly, as in the \texttt{circuit2height} function below.

\vspace{0.25cm}
\begin{verbnobox}[\fontsize{8pt}{8pt}\selectfont]
def circuit2height(qc):
    # get the number of qubits from the circuit
    n = qc.num_qubits
    grid = make_grid(int(2**(n/2)))
    # get the initial state from the circuit
    ket = qi.Statevector(qc.data[0][0].params)
    qc.data.pop(0)
    # evolve this by the rest of the circuit
    ket = ket.evolve(qc)
    # extract the output probabilities
    p = ket.probabilities_dict()
    # determine maximum probs value for rescaling
    max_h = max( p.values() )   
    # set height to rescaled probs value
    height = {}
    for bit string in p:
        if bit string in grid:
            height[grid[bit string]] = p[bit string]/max_h
    return height
\end{verbnobox}
\vspace{0.25cm}

If we were instead to use real quantum hardware, the process would be slightly more involved. Firstly, to obtain the output bit string we would need to measure the qubits. The measurement of all qubits can be done by adding measurements to the end of the circuit using \texttt{qc.measure\_all()}. Each run of the circuit then returns a bit string $b$, drawn from the probability distribution described by the probabilities $p_{b}$. To estimate these probabilities we must then run the circuit for many samples, and count the number of runs that give each output. These results can be expressed as a so-called $\mathtt{counts}$ dictionary, for which $\mathtt{counts[b]}$ represents the number of samples for the bit string $\mathtt{b}$. This is done with

\vspace{0.25cm}
\begin{verbnobox}[\fontsize{8pt}{8pt}\selectfont]
result = execute(qc, backend, shots=shots).result()
counts = result.get_counts()
\end{verbnobox}
\vspace{0.25cm}

where \texttt{backend} describes the device on which the circuit \texttt{qc}, and $\mathtt{shots}$ is the number of samples for which the process is repeated.

By dividing each of the counts values by $\mathtt{shots}$, which specifies the total number of runs, we obtain an estimation of the probabilities. However, since we must rescale these values in any case to regain the original image (to ensure that the maximum value is equal to $1$), we can simply use the counts values directly instead of the probabilities. Note that, as mentioned earlier, the overhead of sampling $\mathtt{shots}$ times means that running on real quantum hardware is less efficient for this specific application.

However we run the process in \texttt{circuit2height}, in combination with \texttt{height2circuit} we can now implement the simple process of encoding and then recovering the example image \texttt{height}.

\vspace{0.25cm}
\begin{verbnobox}[\fontsize{8pt}{8pt}\selectfont]
qc = height2circuit(height)
new_height = circuit2height(qc,backend)
\end{verbnobox}
\vspace{0.25cm}

\begin{figure}[!tbp]
  \centering
  \subfloat[]{\includegraphics[width=0.45\columnwidth]{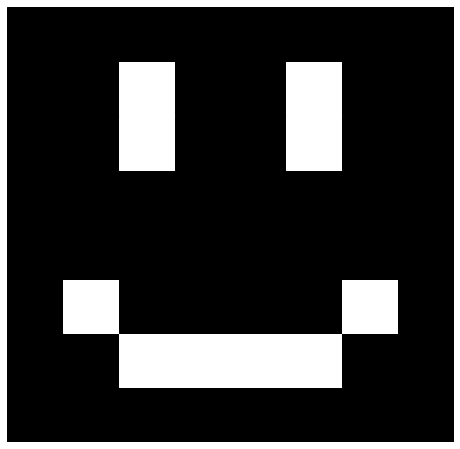}}
      \hfill
  \subfloat[]{\includegraphics[width=0.45\columnwidth]{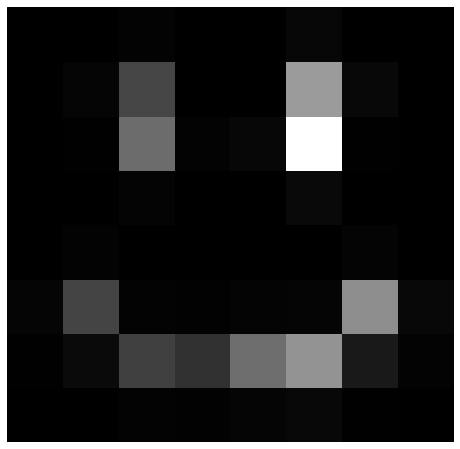}}
  \subfloat[]{\includegraphics[width=0.45\columnwidth]{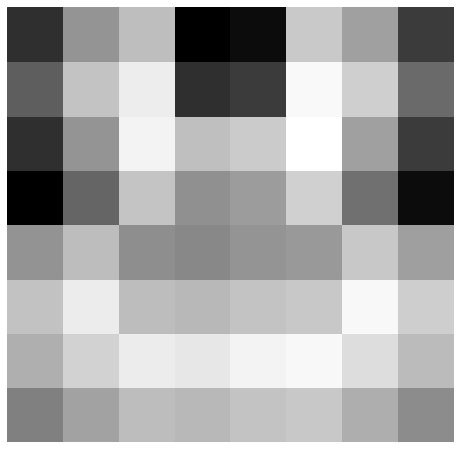}}
  \caption{  \label{face} (a) A simple face encoded in and then recovered from a quantum circuit. (b) The image after an $R_y(\pi/10)$ gate on all qubits. (c) The same image as in (b), but with values plotted logarithmically. }
\end{figure}

More interesting is to manipulate the image by adding quantum gates to the circuit. Though there are many possible choices we could make of what gates to apply, we consider something very straightforward: simply applying $R_y$ to all qubits by a given angle $\theta$. The results are shown in Fig.~\ref{face}. Note that the effect appears very slight, due to the blurring occurring with an exponential distribution. The effect can be seen more clearly when the values are plotted logarithmically.

\begin{figure}[!tbp]
  \centering
  \subfloat[]{\includegraphics[width=0.45\columnwidth]{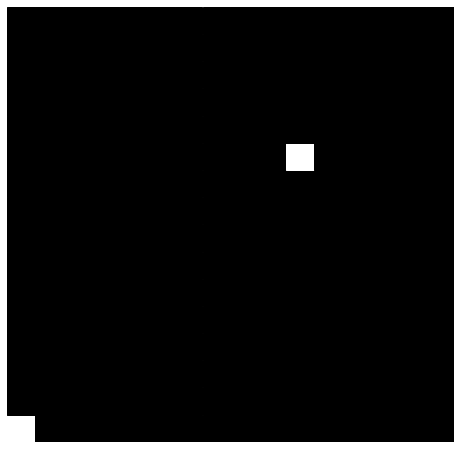}}
  \subfloat[]{\includegraphics[width=0.45\columnwidth]{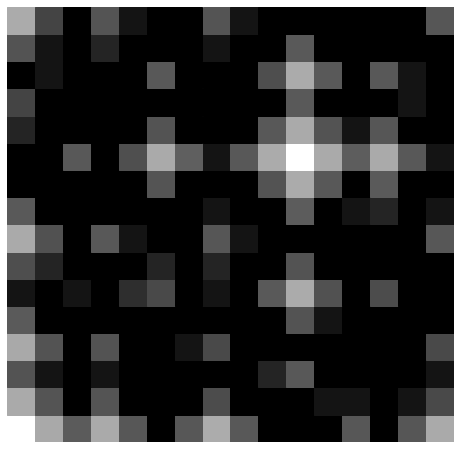}}
    \hfill
  \subfloat[]{\includegraphics[width=0.45\columnwidth]{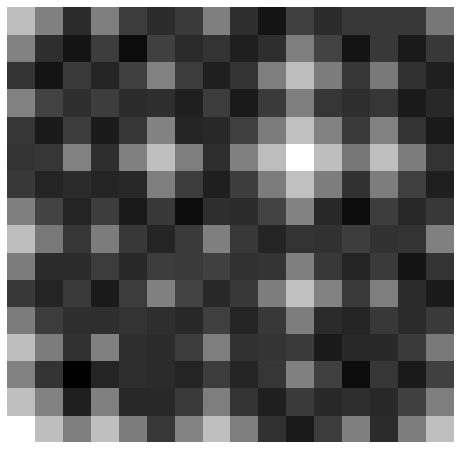}}
  \subfloat[]{\includegraphics[width=0.45\columnwidth]{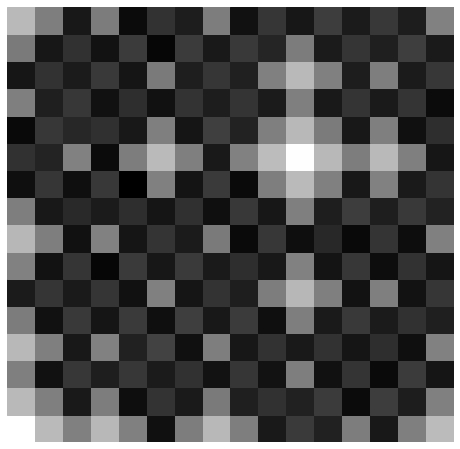}}
    \hfill
  \subfloat[]{\includegraphics[width=0.45\columnwidth]{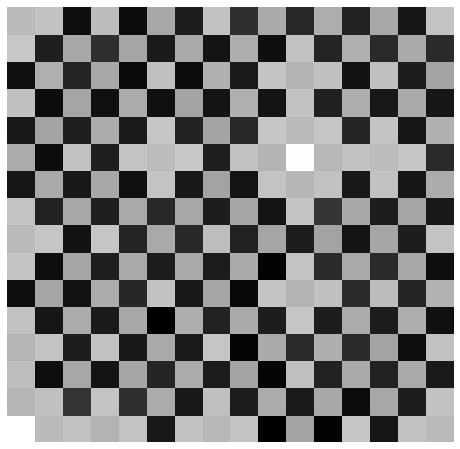}}
  \subfloat[]{\includegraphics[width=0.45\columnwidth]{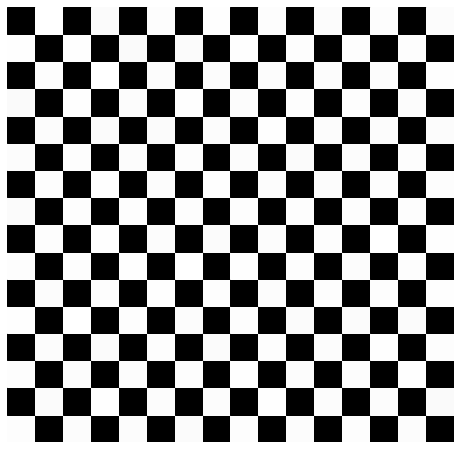}}
  \caption{  \label{ghz} A rotation of the GHZ state by (a) $\theta=0$, (b) $\theta=0.1 \pi$,  (c) $\theta=0.2\pi$,  (d) $\theta=0.3 \pi$, (e) $\theta=0.4 \pi$,  (f) $\theta=0.5 \pi$, all. plotted logarithmically. }
\end{figure}

The fact that the blur effect is due to an interference process can be most easily seen when large angles are used. This is shown in Fig.~\ref{ghz}, where the initial two-pixel images represents the so-called GHZ state~\cite{ghz}.

Here the small angles show a relatively simple blurring effect. However, rather than simply blurring out into a uniform distribution, the interference effects at $\theta=0.5 \pi$ create a checkerboard pattern. Larger angles would see the original points reform, leading to the original image at $\theta=\pi$. The interference pattern at $\theta=0.5 \pi$ depends strongly on the initial state, and so would be different if other pixels were chosen in the original image.

\subsection{Applications of the method}

The quantum blur method presented has been developed and tested during projects for various game jams. It was primarily used to procedurally generate textures from simple seed images. Some of these applications are outlined below.

The first use in a game jam was for PyWeek 27 in which it was used as the basis for an art toy: the `Quantograph'~\cite{pyweek:27}. This allowed users to choose a seed image as well as set of parameters. Multiple quantum circuits based on these parameters were then implemented, each slightly different from the last, to create successive frames for an animation. This provided a visualization of quantum interference effects distorting the seed image. In this project, the method presented in the previous sections was adapted to allow for colour images. This was done simply by running three different quantum processes, with one for each rgb colour channel. Similar animations were also used in a game for the GMTK Game Jam 2019~\cite{gmtk:19}. In this case it served as a transition animation, where the screen was scrambled and unscrambled by the quantum process when transitioning between levels.

\begin{figure}[!tbp]
  \centering
  \subfloat[]{\includegraphics[width=0.4\columnwidth]{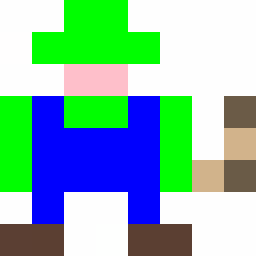}}
  \subfloat[]{\includegraphics[width=0.4\columnwidth]{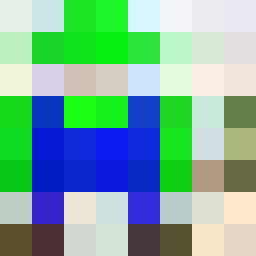}}
    \hfill
  \subfloat[]{\includegraphics[width=0.4\columnwidth]{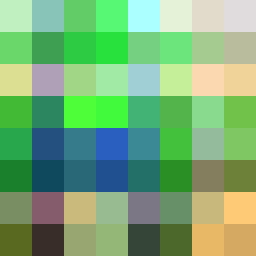}}
  \subfloat[]{\includegraphics[width=0.4\columnwidth]{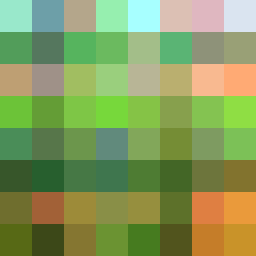}}
    \hfill
  \subfloat[]{\includegraphics[width=0.4\columnwidth]{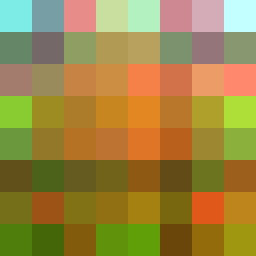}}
  \subfloat[]{\includegraphics[width=0.4\columnwidth]{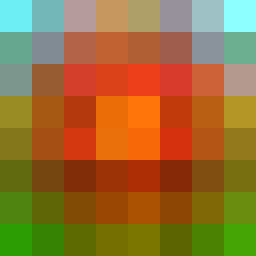}}
    \hfill
  \subfloat[]{\includegraphics[width=0.4\columnwidth]{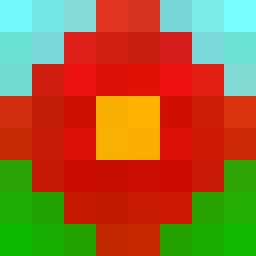}}
  \subfloat[]{\includegraphics[width=0.4\columnwidth]{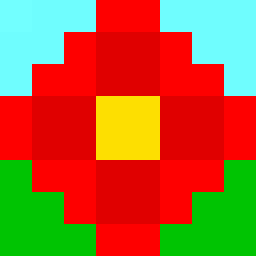}}
  \caption{   \label{teleport} Transition between a plumber and a flower. The small size of these images is not a limitation of the method, but was chosen in line with the theme of a game jam.}
\end{figure}

As an extension of this idea, transitions between two seed images can be achieved using a teleportation-like effect. Specifically using the \texttt{SWAP} gate, which acts on two equal sized registers of qubits, and simply swaps their states. Just as $R_x$ and $R_y$ can be viewed as fractional forms of the \texttt{NOT} gate, so too can we define fractional forms of the \texttt{SWAP} gate. By encoding two different images in different registers of qubits, and then applying different fractions of the \texttt{SWAP} gate to generate images for different frames, we can visualize the intermediate states of this quantum \texttt{SWAP} operation, creating an animation of a teleportation-like effect. An example of this for few-pixel images is shown in Fig.~\ref{teleport}.

\begin{figure}[!tbp]
  \centering
  \subfloat[]{\includegraphics[width=0.45\columnwidth]{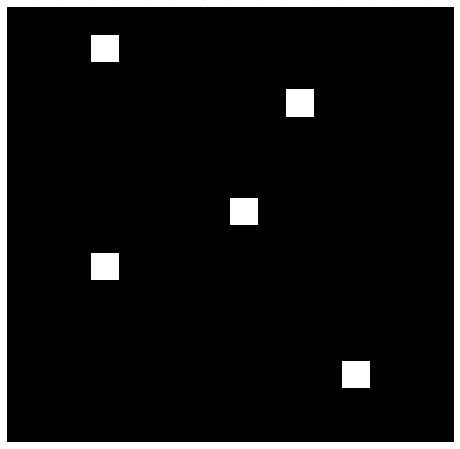}}
   \subfloat[]{\includegraphics[width=0.45\columnwidth]{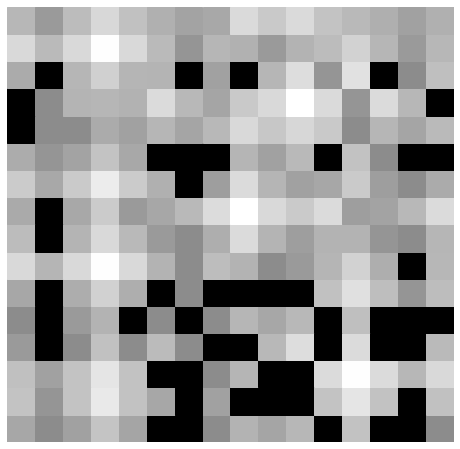}}
    \hfill
  \subfloat[]{\includegraphics[width=0.45\columnwidth]{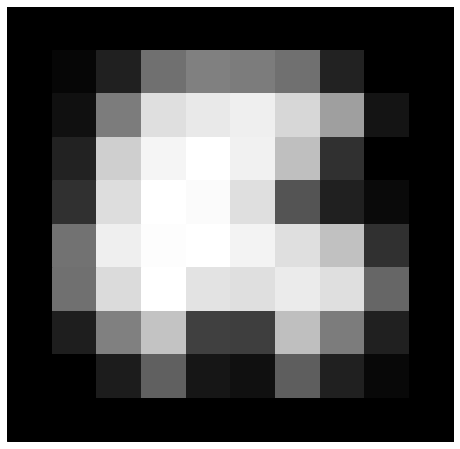}}
  \subfloat[]{\includegraphics[width=0.45\columnwidth]{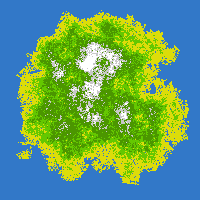}}
  \caption{   \label{island} (a) A randomly chosen $16\times 16$ pixel seed image. (b) The seed image with $\theta=0.15 \pi$, plotted logarithmically. (c) A simple $10\times 10$ pixel layout for an island. (d) A $200 \times 200$ pixel island with texture generated by random placement of the manipulated seed image over the simple island layout.}
\end{figure}

In Ludum Dare 44~\cite{ld:44}, the quantum blur effect was used to generate maps on which a simple game was played. An example of the method used is depicted in Fig.~\ref{island}. Specifically, the $16 \times 16$ patch of texture in Fig.~\ref{island} (b) is created from the randomly generated seed image of Fig.~\ref{island} (a). Hundreds of random variants of this are then created by a shuffling process, in which alternative mappings between bit strings and coordinates are randomly generated and the corresponding images are constructed using the same set of probabilities each time. This generates hundreds of textures while only using the quantum process once, with an entire time taken of less than 10 seconds on a laptop. These textures are then randomly placed onto a $200 \times 200$ pixel image, where the probability of placement at at given point is governed by the height of a predetermined simple layout. The $10\times 10$ pixel layout in Fig.~\ref{island} (c) is used in the example (stretched out to the required size). With this whole process the island of Fig.~\ref{island} (c) was generated, where colours represent the type of terrain at different heights.

\begin{figure}[!tbp]
  \includegraphics[width=\columnwidth]{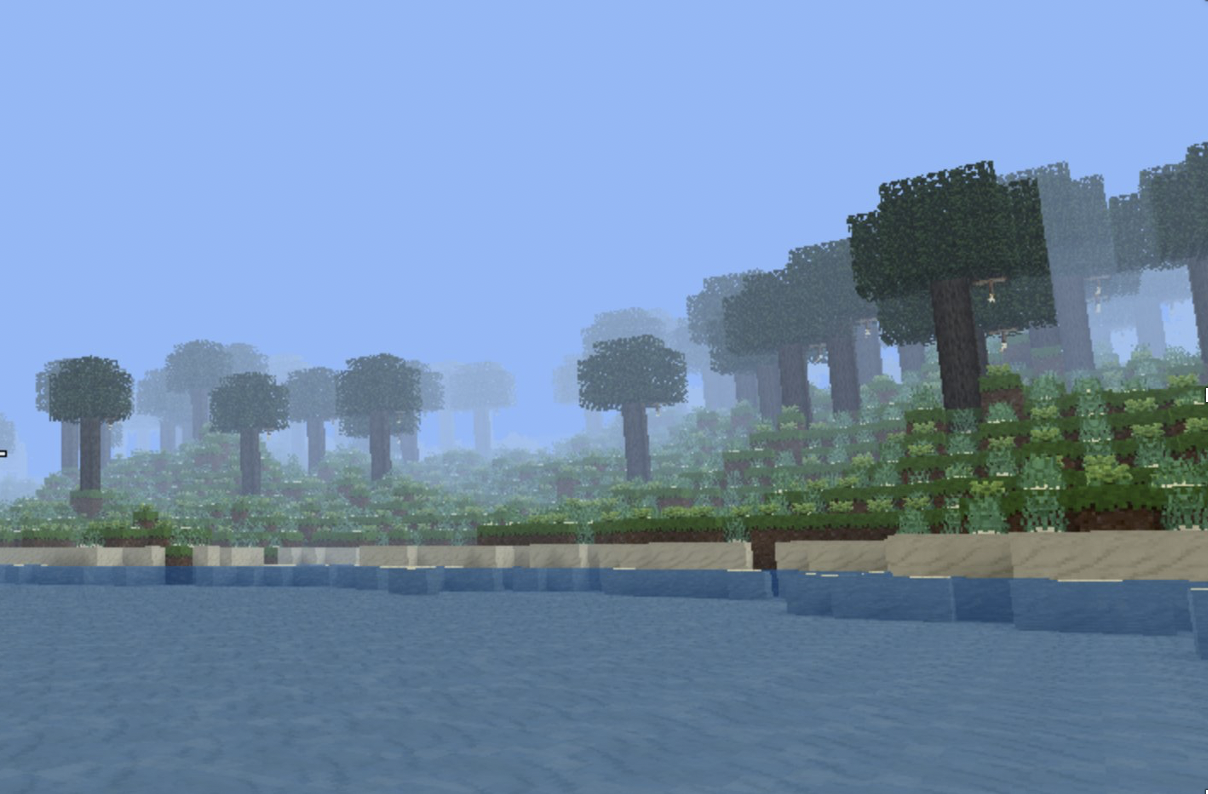}
  \caption{\label{3Disland} An island generated using the quantum blur effect, and rendered in 3D in QiskitBlocks.}
\end{figure}

During PROCJAM 2019~\cite{procjam:19}, the above method was adapted to create islands suitable to be rendered in 3D, as shown in Fig.~\ref{3Disland}. This was done using a voxel based game engine, meaning that the height was rounded down at each point. This left a value $0 \leq r < 1$ at each point as the difference between the true and rounded values. These values were used to decide the position of objects such as trees. This method was subsequently used in the educational game QiskitBlocks~\cite{qiskitblocks}.

The method has also been applied to other forms of discrete data, using the results of \texttt{make\_line} as a starting point for the mapping. For example, we can consider a given musical note to be located in a 3D space consisting of coordinates describing which bar it is in, placement within that bar, and its octave. Each coordinate corresponds to a line of bit strings of suitable length, and the total coordinate corresponds to the combination of these bit strings. With this we can then convert music to and from quantum states, and the `blurring' effect of quantum gates will cause notes to bleed across bars, and between different bars and octaves.

\section{Comparison to other methods}

When comparing to other methods, the most obvious comparison is to the box blur. For an $L \times L$ image, a blur of any radius $r$, can be done with just $O(L^2)$ complexity~\cite{blur}: linear in the number of points.

The quantum blur presented here uses the simulation of $n=  \left\lceil \log_2 (L^2) \right\rceil$ qubits, which requires manipulation of a vector of length $2^n$. The blur effect is provided by performing a rotation gate on all qubits. Doing these individually would result in a total complexity of $O(n 2^n) = O( L^2 \log L)$. This is not significantly worse than the box blur, and methods could be found to improve the complexity. Nevertheless, the quantum method cannot beat the complexity of the standard box blur.

Furthermore, content generated by the quantum method has artefacts that result from quantum interference effects and the way that large images are squeezed into a small number of qubits (primarily the effects of non-neighbouring points having neighbouring bit strings). Such artefacts, which make it obvious that the content was generated algorithmically and which provide noticeable traces of the algorithm used, are usually regarded as being problematic in procedural generation~\cite{backus:17}. Significant efforts are therefore typically made to avoid such effects. 
However, these artefacts might actually prove useful in the quantum case. This is because quantum computing is an area which is of interest to many people, not just because of the results that it will provide but also because of the connection that the hardware and software have to the science of quantum physics. Since quantum physics is a popular topic used (or misused) in science fiction, artefacts which provide a signature of a quantum origin could help to build a desired aesthetic in a sci-fi context, or provide the player with the sense of a more genuine sci-fi experience.

The idea that quantum artefacts may be beneficial for some use cases is inspired by the way that external parties have already been using the method proposed here. Nevertheless, this idea is primarily speculation at this stage. Studies into the public perception of procedurally generated content has so far only been done for standard computational methods~\cite{lamb:18}. However, it might now be timely to perform similar studies into the perception of quantum results.

Nevertheless, it is not the intent in this paper to propose a method suitable solely for sci-fi projects, as the examples shown in the last section should demonstrate. However, this may well be a niche for which quantumly generated content can excel within the near-term.

\section{Conclusions}

For many applications, the usefulness of quantum computing will begin once the hardware is able to implement the thousands of qubits needed for standard textbook algorithms~\cite{shor,grover,qiskit-textbook}. For other applications, a quantum advantage will be found for nearer term devices with hundreds of qubits~\cite{nisq}. For a few applications, however, designing algorithms using the principles of quantum computing could be a useful venture even while those algorithms can be run more easily on conventional computers than quantum hardware. In these cases, the design of immediately useful quantum methods can begin now, and will only grow in usefulness and sophistication as ever more powerful quantum hardware emerges.

The quantum blur effect presented here shows that procedural generation is such a case. It is targeted specifically at using simulators or currently available prototype quantum hardware, and uses quantum operations to create an effect that could be useful for various tasks within procedural generation. It is clearly a very simple tool, and is not expected to become the first `killer app' of the quantum era in itself. Nevertheless, it shows the potential that quantum computing can have even with current computational resources, and will hopefully inspire others to investigate what quantum computing can do for them.

\section*{Acknowledgments}

Thanks to the organizers and participants of the PyWeek, Ludum Dare, GMTK and PROCJAM game jams, at which much of this work was performed.

\bibliographystyle{IEEEtran}
\bibliography{refs}

\end{document}